\documentclass[prl,twocolumn, showpacs,superscriptaddress]{revtex4}
\usepackage{graphicx}
\usepackage{color}
\usepackage{subfigure}
\usepackage{amsmath}

\begin{document}

\title{Optically modulated conduction in chromophore-functionalized
single-wall carbon nanotubes}
\author{J. M. Simmons}
\affiliation{Department of Physics, University of Wisconsin-Madison, Madison, WI 53706}
\author{I. In}
\affiliation{\mbox{Department of Materials Science and Engineering, University of Wisconsin-Madison, Madison, WI 53706}}
\author{V. E. Campbell}
\affiliation{\mbox{Department of Materials Science and Engineering, University of Wisconsin-Madison, Madison, WI 53706}}
\author{T. J. Mark}
\affiliation{\mbox{Department of Materials Science and Engineering, University of Wisconsin-Madison, Madison, WI 53706}}
\author{F. L\'{e}onard}
\affiliation{Sandia National Laboratories, Livermore, CA 94551}
\author{P. Gopalan}
\affiliation{\mbox{Department of Materials Science and Engineering, University of Wisconsin-Madison, Madison, WI 53706}}
\author{M. A. Eriksson}
\affiliation{Department of Physics, University of Wisconsin-Madison, Madison, WI 53706}

\begin{abstract}
We demonstrate an optically active nanotube-hybrid material by
functionalizing single-wall nanotubes with an azo-based chromophore. Upon UV
illumination, the conjugated chromophore undergoes a cis-trans isomerization
leading to a charge redistribution near the nanotube. This charge
redistribution changes the local electrostatic environment, shifting the
threshold voltage and increasing the conductivity of the nanotube
transistor. For a $\sim$1-2\% coverage, we measure a shift in the threshold
voltage of up to 1.2~V. Further, the conductance change is reversible and
repeatable over long periods of time, indicating that the chromophore
functionalized nanotubes are useful for integrated nano-photodetectors.
\end{abstract}

\maketitle


Recent experiments demonstrating the use of individual single-wall carbon
nanotube transistors as both photoabsorbers \cite{BalasubNanoLett,FreitagNanoLett,OhnoJJAP} and emitters \cite{ChenScienceIBM,MisewichScience} have enhanced the prospects for using
nanotubes in optoelectronic devices. In the case of photoabsorbers, the
ability to tune the absorption window is limited by difficulties in
synthetically controlling the nanotube chirality, and thus the band
structure, as well as the formation of bound excitons \cite{KaneMelePRL,WangExciton}. In principle these problems can be overcome by using chromophore functionalized nanotubes \cite{GuoJACSColumbia,HechtNanoLett,ValentiniJAP} where the molecular photoabsorption can be synthetically tuned independent of the nanotube
electronic structure. However, there are several issues that need to be
addressed to make this approach viable: interactions between the chromophore
and the nanotube may lead to scattering of the charge carriers and device
degradation; in the case of photoswitching by charge transfer between the
chromophore and the nanotube, the hybrid must have the correct band
alignment between the molecular orbitals and the nanotube electronic states;
photobleaching mechanisms in many chromophore systems can lead to the loss
of optical activity.

To circumvent these issues, we demonstrate an approach based on the use of
photoisomerization of organic chromophores to optically modulate
the conduction in individual nanotube transistors. Chromophore isomerization causes a significant change in the molecular dipole moment, effecting a change in the local electrostatic environment.  Atomistic modeling of the nanotube
transistor in the presence of dipoles is used to support this
mechanism. These optically active nanotube-hybrids show no indication of
bleaching and can be reliably switched over long periods of time. In
contrast to a charge-transfer effect, the dipolar switching mechanism does
not rely on the specific electronic states of the nanotube channel; thus
this mechanism is general and can be applied to other semiconducting
nanomaterial systems. In addition, synthetic control of the chromophore
offers the ability to tune both the absorption wavelength and sensitivity of
the nanotube-hybrid transistor.

\begin{figure}[h]
\centering \includegraphics[width=3.in]{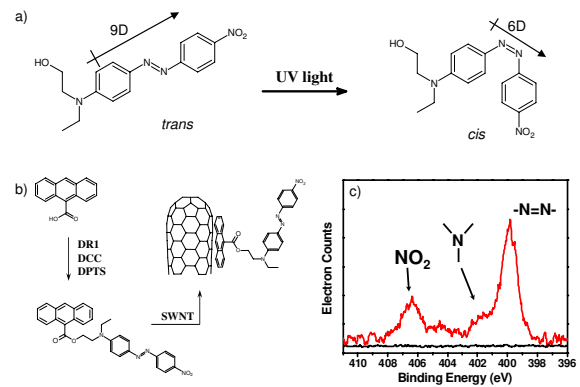}
\caption{(a) Under UV light, DR1 isomerizes from the equilibrium \emph{trans}
conformation to the metastable \emph{cis} conformation. In doing so, the
molecular dipole moment changes from 9~D to 6~D. (b) Reaction scheme for
non-covalent nanotube functionalization. 
(DPTS=4-(dimethylamino)-pyridinium \emph{p}-toluenesulfonate) (c) N(1s)
core level photoelectron spectrum of the physisorbed anthracene-DR1
functionality.}
\label{DR1switch}
\end{figure}

We use an azobenzene-based chromophore, Disperse Red~1 (DR1), as the
photo-switching side-wall functionality. This molecule is known
to isomerize under UV illumination, accompanied by a significant change in
the dipole moment \cite{AtassiDR1Moment}. Experiments
have shown that at equilibrium, the DR1 chromophore is in the \emph{trans}
conformational state (Fig.~\ref{DR1switch}) in which there is significant
orbital overlap between the phenyl rings.
Due to this conjugation and the presence of the strongly electronegative
nitro end group, the \emph{trans} conformation has a considerable $\sim $9~D
dipole moment \cite{AtassiDR1Moment}. Under UV light, the chromophore
isomerizes to the \emph{cis} conformation (Fig.~\ref{DR1switch}) where the orbital overlap is significantly
reduced, leading to a smaller $\sim $6~D dipole moment. If left in ambient
conditions, the \emph{cis} isomer will relax to the more stable and less
sterically hindered \emph{trans} conformation.

We functionalize nanotubes using non-covalent attachment of the DR1
chromophore via an anthracene tether. Though such chromophores can be
attached to the nanotube side-walls either covalently or non-covalently,
non-covalent attachment is advantageous because it only weakly perturbs the
nanotube electronic states. Polycyclic aromatic hydrocarbons such as pyrene
and anthracene have been shown to physisorb onto nanotubes by forming $\pi
-\pi $ bonds to the nanotube side-wall with little charge transfer \cite{ChenJACSDai,LuJACS}. 
 Since most $\pi -\pi $ bound molecules can be
easily removed using common solvents, non-covalent attachment also offers
the ability to reversibly functionalize nanotubes. The anthracene modified
DR1 is synthesized from 9-anthracenecarboxylic acid and DR1 using a
dicyclohexylcarbodiimide (DCC) esterification reaction (Fig.~\ref{DR1switch}%
). The crude product is purified by silica gel chromatography and the
structure is confirmed using $^{1}$H-NMR \cite{HNMR}. After purification,
the anthracene-DR1 is dissolved in DMF for application to the nanotubes.
Pure anthracenecarboxylic acid is also dissolved for use as a non-functional
control.

The individual nanotube transistors are fabricated on highly doped silicon
wafers with a 500~nm thermal oxide using chemical vapor deposition (CVD) and
subsequent lithography. The nanotubes are grown using iron and molybdenum
nanoparticle catalysts at 900~$^\text{o}$C using methane feedstock with
hydrogen co-flow \cite{KongCPLCVD}. Under these CVD conditions, the average
nanotube diameter is $\sim$1.6~nm. After growth, electron and atomic force
microscopy are used to locate individual nanotubes and electron beam
lithography, metal deposition and 400~$^\text{o}$C forming gas anneal (4:1 Ar/H$_2$) are used to form contacts. A total
of 12 devices are processed for this study, nine of which functioned as transistors after fabrication.

After annealing, a drop of the anthracene-DR1 (or anthracenecarboxylic acid)
solution is placed onto the chip and the sample is washed to remove the
non-specifically bound chromophores. X-ray photoelectron spectroscopy (XPS, Fig.~\ref{DR1switch}c)
measurements confirm that the chromophore remains after
washing and yield a coverage of 1-2 molecules per 100 nanotube carbon atoms, calculated from the N(1s) to C(1s) intensity ratio \cite{XPShandbook}. I-V$_{g}$ characteristics
of individual single-wall carbon nanotube transistors before and after
adsorption of the chromophore show minimal changes in the drain current,
indicating that the anthracene-DR1 molecules cause minimal scattering of the
charge carriers, and no degradation in device performance. Devices can
exhibit threshold voltage shifts due to small chirality-dependent charge
transfer from anthracene \cite{LuJACS}. Gate voltage scans to high positive
voltages show no indication of the n-channel opening up, indicating that the
transistors are unipolar.

\begin{figure}[h]
\centering \includegraphics[height=1.75in]{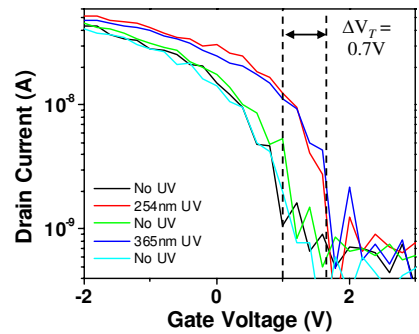}
\caption{Transistor characteristics showing threshold voltage shifts under
UV light. The threshold voltage is shifted by 0.7~V for both 254~nm and
365~nm light and is fully reversible.}
\label{IVgtraces}
\end{figure}

UV-induced switching is performed using a handheld UV lamp with 254~nm and
365~nm lines and a low intensity of $\sim $100~$\mu $W/cm$^{2}$. Electrical
measurements are performed in a nitrogen purged cell to eliminate the
potential for ozone oxidation from UV excitation of the atmospheric oxygen %
\cite{SimmonsJPCB}. While measuring the drain current through a
nanotube transistor, the gate voltage is varied to acquire a series of I-V$%
_{g}$ characteristics (Fig.~\ref{IVgtraces}). Before illumination with UV
light, the transistor shows p-type behavior with a threshold voltage of $%
\sim $1~V. When the chromophore is isomerized to the \emph{cis}
conformation, the threshold voltage is shifted to the right, in this case by
0.7~V, and does not depend on the wavelength of UV light used.
The threshold voltage shift under illumination was measured for five nanotube transistors; all show
positive threshold shifts with a range of 0.6-1.2~V. This shift could
indicate a charge transfer mechanism \cite{HechtNanoLett} or a change in the
local electrostatic environment \cite{MarcusJAP,LarrimoreNanoLett}. 
Charge transfer from the chromophore to the nanotube is inhibited by the
alkane spacer and anthracene tether separating the chromophore and the nanotube and control
experiments show no
threshold voltage shift, indicating that there is no photo-induced charge
transfer from the anthracene tether. An estimate of the amount of
charge transfer needed for a 1V threshold voltage shift gives
0.07e/molecule, so we cannot entirely rule out charge transfer.  However, since the UV photoisomerization and concurrent dipole moment change of DR1 is well established, we propose that the
dipole moment change acts as a small local negative gate voltage. The large
transistor threshold voltage swing is due to the relatively short spacer
group used to separate the chromophore from the nanotube ($\sim $%
1~nm for anthracene-DR1).

\begin{figure}[h]
\centering \includegraphics[width=2.5in]{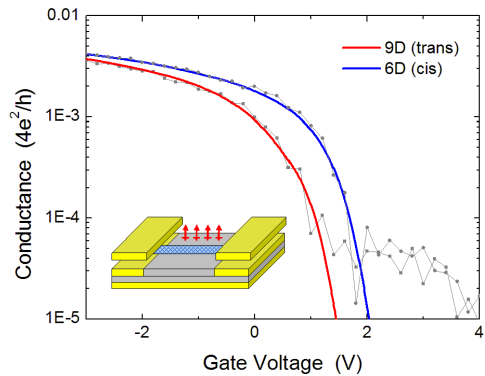}
\caption{Device simulations of the nanotube conductivity for both
chromophore isomers (solid lines) compared with the experimental data
(gray symbols). The transistor characteristic for the 9~D \emph{trans} isomer (red
curve, left) shifts toward positive gate voltages when the chromophore switches to
the 6~D \emph{cis} isomer (blue curve, right). As seen in the experiments, the
threshold voltage is shifted to the right by 700~mV. Inset: Sketch of nanotube transistor used in the simulations, with the red arrows indicating the molecular dipoles.}
\label{FrancoisConductance}
\end{figure}

To further test this mechanism, we performed quantum transport calculations
of the nanotube transistor characteristics in the presence of dipoles. In brief, we use the non-equilibrium Green's function formalism implemented in a tight-binding formalism in conjunction with a Poisson
solver to self-consistently calculate the charge and the electrostatic
potential in the device, and the conductance \cite{Francois1,Francois2}
 (this method is applicable to calculations of steady-state
quantities; the time-dependence of the switching, as discussed below, is
beyond the scope of this approach). A schematic of the simulated device is
shown in Fig.~\ref{FrancoisConductance}; it consists of a (17,0) nanotube
(bandgap 0.55 eV, radius 0.66 nm) sitting on SiO$_{2}$. At the ends of the
computational cell, the nanotube is sandwiched between two metallic plates,
forming the source and drain contacts. The bottom and top metallic plates
have thicknesses of 2.25 nm and 2 nm, respectively, and are separated from
the nanotube by 0.3 nm. The metal work function is 1 eV below the nanotube
midgap, and after self-consistency, sits slightly below the valence band
edge, creating an ohmic contact. Because of the computational demands of the
calculations, we cannot simulate the actual experimental device; instead, we
focus on a smaller device of 35 nm channel length and with a gate oxide
thickness of 4.5 nm. The results of our computations are then related to the
experimental device by scaling the gate voltage by the ratio of the
capacitances of the experimental and simulated device, using the expression
for the capacitance $2\pi \varepsilon /\ln (2t/R)$ where $t$ is the gate
oxide thickness and $R$ is the nanotube radius. Because of the high density
of molecules on the nanotube surface, we assume that the dipole moment is
perpendicular to the nanotube surface, and is modeled by a positive
(negative) point charge of magnitude equal to the electron charge and at
distance $h$ ($h+d)$ above the nanotube. From the molecule configuration we
estimate $h=1$~nm, and use a three-dimensional Gaussian distribution of the
charge. The dipoles are evenly distributed on the nanotube using the range
of experimentally measured chromophore densities. The \emph{trans} to \emph{%
cis} isomerization is studied by changing the value of $d$ from 0.1 to
0.067~nm, reproducing the molecular dipole moments of 9 D and 6 D.

The calculated conductance as a function of gate voltage is shown in Fig.~%
\ref{FrancoisConductance} for the \emph{trans} and \emph{cis} cases,
overlaid on the experimental data.
A dipole spacing of 0.35~nm is used, corresponding to the midpoint of the
range of chromophore densities measured by XPS. (The calculated
conductance was rescaled by a factor 1/140 to match the experimental
conductance in the ON state of the nanotube transistor. The lower
conductance of the experimental devices may be due to scattering in the
nanotube or the presence of a small Schottky barrier at the contacts. The
numerical results were also shifted by 0.65 V to match the threshold voltage
of the \emph{trans} experimental data. The need for this shift may be due to
doping of the nanotube. The rescaling of the conductance and shift
of the gate voltage are only necessary to reproduce the conductance curve prior to illumination. Once this is obtained, no further
adjustment is made, and the shifted conductance curve is obtained by
changing only the dipole moment.) A threshold voltage shift of 700~mV is
found, in excellent agreement with the experiments and supporting the dipole
change mechanism as the reason for the modulation of the nanotube
conductance. Calculations were performed for several dipole densities. From
these results, the uncertainty in the measured dipole density yields a range
in the predicted threshold shift of approximately 0.45-0.95~V, correlating
well with the 0.6-1.2~V shifts measured for other devices.

We turn next to the kinetics of the switching event, shown in Fig.~\ref%
{AnthDR1timetrace}. In (a), \emph{trans} to \emph{cis} isomerization under
254~nm UV light leads to a uniform increase in the drain current. In
addition, the magnitude of the current change is dependent on the gate
voltage, with the largest signal arising from the subthreshold region of the
I-V$_g$ characteristic near V$_g$=0~V (see Fig.~\ref{IVgtraces}). 
\begin{figure}[h]
\centering \includegraphics[width=2.5in]{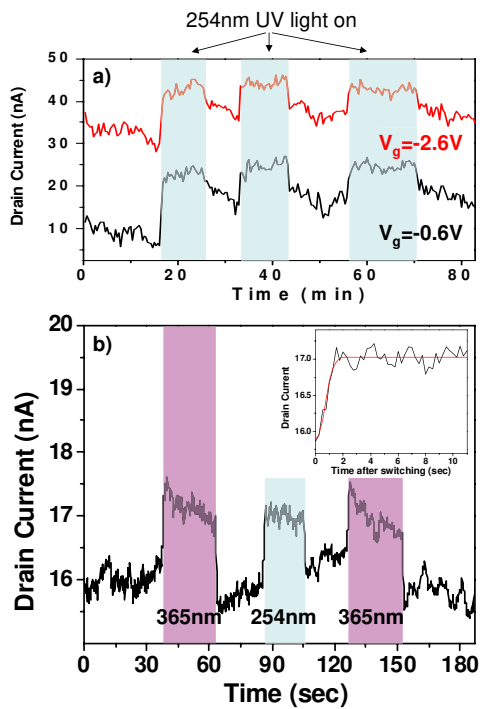}
\caption{Time traces at fixed gate voltage for two different nanotube-hybrid
transistors showing multiple repeatable switching events under both 254~nm
and 365~nm light. Lower inset: Close up view of the transition under 254~nm
light from the bottom graph. The smooth red line is a fit to the data using the
Avrami equation (see text) with a time constant of $\sim $2~s.}
\label{AnthDR1timetrace}
\end{figure}
In both cases shown, the sample was illuminated three times, each of which
led to an appreciable increase in the measured drain current. All fabricated transistors show increased drain current under UV illumination, with abrupt and repeatable transitions. The transistors have shown repeatable switching for more than 100 cycles and over several months.

DR1 has a finite absorption window in the UV region such that most UV
wavelengths will isomerize the chromophore. Specifically, DR1 is sensitive
to both 254~nm and 365~nm UV light, with an
approximate 5:1 absorbance ratio (254~nm:365~nm).
Indeed, we see in Fig.~\ref{AnthDR1timetrace}b that 365~nm UV light also
leads to modulation of the nanotube conduction, though the magnitude of
change is the same for both wavelengths. Given the 5:1 absorbance contrast,
the equal modulation for both UV wavelengths indicates that the
isomerization has saturated under the low intensity UV lamp. Though
the current in this sample decreases under continued 365~nm illumination,
implying a relaxation of some chromophores to the \emph{trans} state, other
samples do not show this decrease and is thus specific to this device.

The inset to Fig.~\ref{AnthDR1timetrace}b shows a close up view of the
transition region under 254~nm UV light. The line through the data is a fit
using the Avrami equation, $\Delta I=1-e^{-Kt^n}$, where $n$=2 \cite%
{AvramiJCP}. Using the time it takes for the current to change from 10\% to
90\% of the full signal, we can extract a rise time of $\sim$2 seconds. The
off-time is approximately the same, as are the switching times for 365~nm
light. This rise time seems surprisingly long for a molecular switching
event, but is consistent with solid state implementations of DR1 in polymer
matrices where molecule-molecule interactions can inhibit the kinetics
of the isomerization \cite{LoucifChemMater}. Because the chromophores form a
dense layer on the nanotube sidewalls, steric interactions between
chromophore units could reduce the overall switching rates. Indeed, the $n$%
=2 Avrami exponent is characteristic of a 2-D growth mechanism from randomly
dispersed nuclei, suggesting that the isomerization proceeds from a few
\emph{cis} conformers in the DR1 layer. The measured switching time is also
as fast or faster than other chromophore-nanotube systems. Experiments using
other chromophore systems show switching times on the order of tens to thousands of seconds \cite%
{ValentiniJAP,GuoJACSColumbia,HechtNanoLett}. Assuming that sterics dominate
the transition rate, diluting the chromophore concentration on the nanotube
sidewall could accelerate the switching.

In summary, we have demonstrated the ability to non-covalently functionalize
individual nanotube transistors using an azobenzene dye. When the
chromophore photoisomerizes, the resulting change in dipole moment modifies
the local electrostatic potential and modulates the transistor conductance
by shifting the threshold voltage. The functionalized transistors show
repeatable switching for many cycles and a modest $\sim $2~s switching time.
The low ($\sim $100~$\mu $W/cm$^{2}$) intensities necessary to optically
modulate the transistor is in stark contrast to measurements of intrinsic
nanotube photoconductivity which typically require 1~kW/cm$^{2}$ intensities %
\cite{FreitagNanoLett}. In addition, synthetic control of the nanotube
hybrid systems should allow the tuning of the absorption window and
magnitude of dipole switching.

\begin{acknowledgments}
The authors thank Sean Cullen for experimental assistance and Bob Hamers for
discussions. This work is supported by the NSF CAREER \mbox{(DMR-0094063)}, MRSEC \mbox{(DMR-0520527)}, and NSEC \mbox{(DMR-0425880)} programs. Sandia is a multiprogram laboratory
operated by Sandia Corporation, a Lockheed Martin Company, for the United
States Department of Energy under contract DE-AC01-94-AL85000.
This work was performed in part at the U.S. Department of Energy, Center
for Integrated Nanotechnologies, at Los Alamos National Laboratory (Contract
DE-AC52-06NA25396) and Sandia National Laboratories (Contract
DE-AC04-94AL85000).
\end{acknowledgments}

\end{document}